\documentclass[aps,prl,twocolumn,english,showpacs]{revtex4-1}
\usepackage{graphicx}
\usepackage{amssymb}
\usepackage{amsmath, amstext, amssymb, amsfonts, amsxtra, mathtools, stmaryrd} 
\usepackage{braket}
\usepackage{textcomp}
\usepackage{relsize}
\usepackage[usenames,dvipsnames]{color}

\usepackage{units}
\usepackage[normalem]{ulem}
\newcommand{\aver}[1]{\langle #1 \rangle}

\begin{document}
\title{Ultracold fermions in a cavity-induced artificial magnetic field}

\author{Corinna Kollath$^1$,Ferdinand Brennecke$^2$}

\affiliation{$^1$HISKP, University of Bonn, Nussallee 14-16, 53115 Bonn, Germany\\
$^2$ Physikalisches Institut, University of Bonn, Wegelerstr.~8, 53115 Bonn, Germany}

\begin{abstract}
 We show how a fermionic quantum gas in an optical lattice and coupled to the field of an optical cavity can self-organize into a state in which the spontaneously emerging cavity field amplitude induces an artificial magnetic field. The fermions form either a chiral insulator or a chiral liquid carrying edge currents. The feedback mechanism via the cavity field enables robust and fast switching of the edge currents and the cavity output can be employed for non-destructive measurements of the atomic dynamics.
\end{abstract}

\date{\today}
\maketitle

The controlled generation of topologically non-trivial quantum phases is of greatest interest since they possess special properties such as extended edge states that can be well protected against destructive environmental effects\cite{XiaoNiu2010}. Thus, materials with topological non-trivial properties have the prospect to be utilized for quantum devices and quantum computing. Well known are topological insulators with protected edge currents created in quantum Hall devices \cite{vonKlitzing1986} by strong magnetic fields. 

For neutral atoms strong {\it artificial} magnetic fields can be created \cite{DalibardOehberg2011} which act similarly as magnetic fields for charged particles. Such artificial magnetic fields have been used in quantum gases confined to optical lattices to realize two showcase models of topologically insulating phases, the Hofstadter model in two-dimensions \cite{JakschZoller2003, AidelsburgerBloch2013,MiyakeKetterle2013,AidelsburgerGoldman2014} or on a ladder geometry \cite{AtalaBloch2014} and the Haldane model \cite{JotsuEsslinger2014}. 

Yet, the dynamic control and detection of topologically non-trivial quantum states remains a great challenge. In order to overcome this difficulty, in this work the dynamical feedback between atoms and an optical cavity is employed to reach a self-organization of topologically non-trivial phases. One fascinating example for a self-organization of a coupled atom-cavity system has been realized recently by placing a bosonic quantum gas into an optical high-finesse resonator subjected to a perpendicular off-resonant pump beam \cite{DomokosRitsch2002,NagyDomokos2008,BaumannEsslinger2010,RitschEsslinger2013,PiazzaZwerger2013}. Above a critical pump strength, the occupation of the cavity mode is stabilized and the bosonic atoms organize into a checkerboard density pattern \cite{BaumannEsslinger2010}. Many different proposals have been put forward to realize the self-organization of more complex quantum phases \cite{RitschEsslinger2013} reaching from the Mott-insulator \cite{LarsonLewenstein2008} over fermionic phases \cite{LarsonLewenstein2008b,MuellerSachdev2012,PiazzaStrack2014,KeelingSimons2014,ChenZhai2014} and disordered structures \cite{StrackSachdev2011,GopalakrishnanGoldbart2011,HabibianMorigi2013,JanotRosenow2013} to phases with spin-orbit coupling \cite{DengYi2014,DongPu2014,PanGuo2014}. 

\begin{figure}
\includegraphics[width=0.6\linewidth]{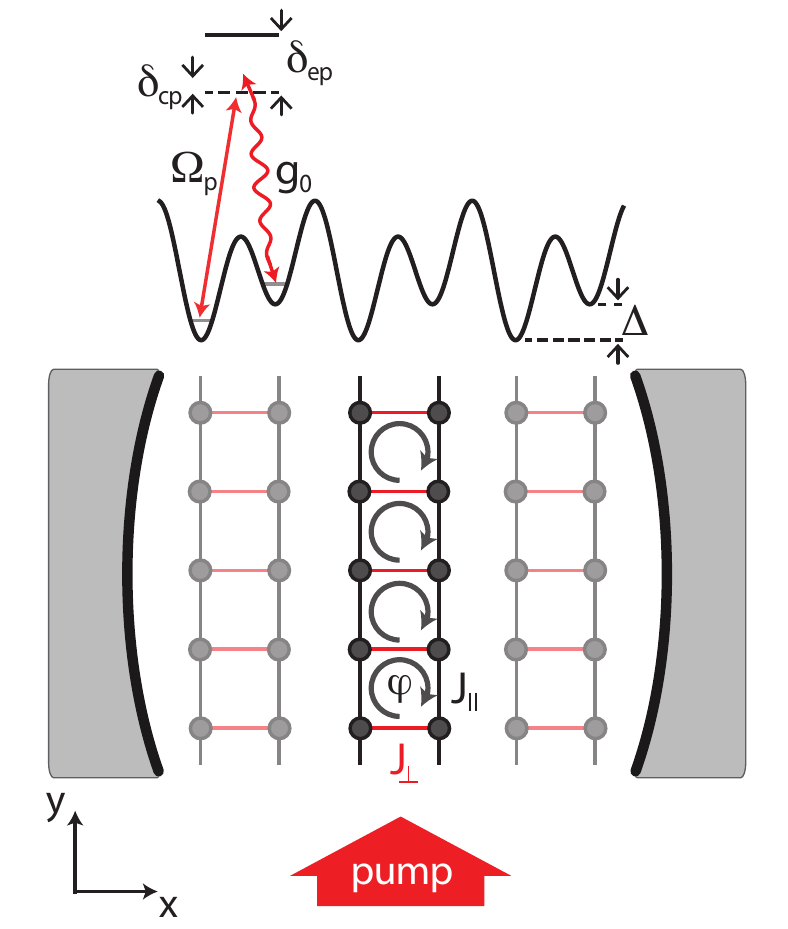}
\caption{\label{fig:setup}(color online) Sketch of the setup. Fermionic atoms in an optical cavity are subjected to an optical lattice potential which creates an  array of ladders (see lower part) for which the tunneling amplitude along the legs is $J_\parallel$. The tunneling along the rungs is strongly suppressed by an potential offset $\Delta$ between neighboring wells. The tunneling along the rung of the ladders is restored by a Raman process using a transverse pump beam and the cavity mode (see upper part).}
\end{figure}

In this work, we engineer a direct coupling mechanism of the cavity photons to the tunneling of atoms in an optical lattice. This is achieved using a Raman transition induced by the combination of a transverse pump beam and the cavity field (Fig.~\ref{fig:setup}). The frequencies of the pump beam and the cavity mode are chosen such that the energy transfer is close to the energy offset between neighboring lattice sites. The resulting process represents an effective tunneling of the atoms. Due to the running-wave nature of the pump beam, a phase $\varphi$ can be imprinted onto the tunneling of the atoms around a plaquette in the presence of cavity photons. We demonstrate that due to this novel coupling mechanism a self-organization into a state can take place in which the cavity mode is occupied and the atoms exhibit protected edge currents. Focusing on a situation in which the optical lattice is used to create an array of decoupled ladders (Fig.~\ref{fig:setup}), either a chiral insulator or a chiral liquid are formed carrying chiral edge currents. Interestingly, the self-organization can set in depending on the filling either suddenly above a critical value of the pump intensity or can be exponentially activated with the pump intensity. In contrast to the application of two classical Raman beams \cite{DalibardOehberg2011}, here the formation of the state relies on the dynamical feedback between the cavity field and the collective atomic tunneling.  The cavity mode reaches rapidly its quasi-stationary state and by its extended nature could accelerate the formation of correlations over the fermionic gas. Furthermore, the cavity output field can be employed for a non-destructive measurement of the tunneling of the atoms.

Cold spin-polarized fermionic atoms are placed into a high-finesse optical cavity which is oriented along the $x$-direction (Fig.~\ref{fig:setup}). We consider a single standing-wave cavity mode with frequency $\omega_c$, since the remaining modes are far detuned and do not contribute to the dynamics. Additionally, a strong lattice potential is applied along the $z$-direction which confines the atoms into decoupled planes. Within these planes, a superlattice structure is created using standing-wave laser beams with wavelength $\lambda_x$ and $2\lambda_x$ along the $x$-direction and  $\lambda_y$ along the  $y$-direction. The phase difference of the two beams along the $x$-direction is chosen such that double wells with a potential offset $\Delta$ are created. The resulting geometry is an array of decoupled ladder structures (Fig.~\ref{fig:setup}) \cite{AtalaBloch2014}. The atoms can tunnel along the leg direction of the ladder. In contrast, the tunneling process along the rungs is strongly suppressed by the potential offset between the two wells. 
To induce a cavity-assisted tunneling process along the rungs, an additional running-wave pump laser of wave length $\lambda_p$ and frequency $\omega_p$ is applied parallel to the legs of the ladders. This is detuned from the dispersively shifted cavity resonance frequency $\tilde{\omega}_c$ by $\hbar(\omega_p-\tilde{\omega}_c)\approx \Delta$ and restores via a near-resonant Raman process the tunneling along the rungs of the ladder geometry (see Fig.~\ref{fig:setup}).

The cavity field and the pump laser are chosen sufficiently far-detuned from the atomic transition frequency $\omega_e$ such that the excited state is only weakly occupied and can be adiabatically eliminated. Thus, an effective description for the atoms in the internal ground state and the cavity mode can be derived (cf.~Refs.~\cite{MaschlerRitsch2008, RitschEsslinger2013}). Performing additionally an expansion of the atomic field operator into Wannier states assuming that the optical lattices are chosen sufficiently strong and neglecting off-resonant terms, the resulting effective Hamiltonian (only a single ladder is considered) reads $H=H_c+H_\parallel+H_{ac}$ with $H_c=\hbar \delta_{cp} a^\dagger a $,
\begin{eqnarray}
H_\parallel&&=-J_\parallel \sum_{j,m=0,1} \left( c^\dagger_{m,j} c_{m,j+1}+\mathrm{H.c.}\right)\nonumber\\
H_{ac}&&= - \hbar\tilde{\Omega} \sum_j \left(a^\dagger e^{i \varphi j} c^\dagger_{0,j} c_{1,j}+\mathrm{H.c.} \right )\nonumber
\end{eqnarray}
Here, $c_{m,j}$ are the annihilation operators of an atom residing on leg $m=0,1$ and site $j$. The term $H_\parallel$ describes the tunneling of the atoms along the legs of the ladder with amplitude $J_\parallel$. The operator $a$ is the annihilation operator of a cavity photon in a frame rotating at frequency $\omega_p-\Delta/\hbar$, and $H_c$ is the corresponding bare Hamiltonian. The two-photon detuning from the resonant tunneling transition between neighboring sites of a rung is denoted by $\delta_{cp}=\left[\tilde{\omega}_c-\omega_p+\Delta/\hbar \right]$, and the AC-Stark shift caused by intra-cavity photons is neglected. The term $H_{ac}$ describes the cavity-induced tunneling of the atoms along the rungs. Its amplitude is given by $\hbar\tilde{\Omega}=\frac{\hbar\Omega_p g_0}{\omega_e-\omega_p}\phi_{_\parallel}\phi_{\perp}$, where $\Omega_p$ denotes the Rabi frequency of the pump and $g_0$ the vacuum-Rabi frequency of the cavity. The factors $\phi_\parallel$ and $\phi_{\perp}$ are effective parameters which can be tuned via the geometry of the optical lattice and the cavity mode \cite{supp}.

Due to the momentum transfer in the cavity-assisted Raman transition a site dependent phase $\varphi j$ is transferred to the atoms when they tunnel along the rungs. Thus, in the presence of a finite cavity field amplitude the fermions feel an artificial magnetic field oriented perpendicular to the $x$-$y$ plane. The amplitude of this magnetic field is determined by the wave vector of the running-wave pump beam. For definiteness we choose $\lambda_p=2\lambda_y$ which results in a flux of $\varphi=\pi/2$ enclosed by the plaquettes of the ladders. Other values of $\varphi$ are also realizable. In contrast to previous setups \cite{JakschZoller2003, AidelsburgerBloch2013,MiyakeKetterle2013,AidelsburgerGoldman2014}, here the artificial magnetic field is not a static field imprinted by external laser beams, but dynamically emerges due to off-resonant scattering of pump light into the cavity mode. The resulting feedback mechanism between the rung tunneling process and the cavity field population leads to a self-organization of the system into a state in which the atoms are subjected to a strong artificial magnetic field. We will analyze the arising states in more detail in the following using a mean-field description for the cavity mode.

Assuming that the cavity field reaches its steady state value rapidly compared to the timescale of the atomic motion, the expectation value  $\alpha=\aver{a}$ can be approximated by the expression $\alpha= \frac{\tilde{\Omega}}{(\delta_{cp}-i \kappa)}\aver{  K_\perp}$, where $\kappa$ is the damping rate of the cavity field. This means that on average the cavity field amplitude is proportional to the directed tunneling $K_\perp =\sum_j  e^{i \varphi j}  c^\dagger_{0,j} c_{1,j}$ of the fermions on all rungs. Note, that the complex phase of $\alpha$ can be chosen freely which reflects the $U(1)-$symmetry of the problem. This will be spontaneously broken in an experiment, and in the following we choose without loss of generality $\aver{K_\perp}>0$. 

The average value $\alpha$ of the cavity field can be substituted into the equations of motion for the fermionic operators. The resulting fermionic evolution can be described by the effective Hamiltonian 
\begin{eqnarray}
H_{F}=H_\parallel - (J_\perp K_\perp+\mathrm{H.c.}).
\end{eqnarray}

The parameter $J_\perp= A \aver{K_\perp^\dagger}$ with $A= \frac{\hbar \tilde{\Omega}^2 \delta_{cp}}{\delta_{cp}^2+ \kappa^2}$ has to be determined self-consistently, because the expectation value of the directed tunneling $\aver{K_\perp}$ does itself depend on the value of $J_\perp$. A self-consistent non-zero solution only exists for a detuning $\delta_{cp}>0$, since only for a positive  value of $J_\perp$ the expectation value of the directed tunneling  $\aver{K_\perp}$ is positive. A non-vanishing solution of $J_\perp$ implies a finite expectation value of the photon operator by the relation $\alpha= J_\perp(\delta_{cp}+i \kappa)/(\hbar \tilde{\Omega} \delta_{cp}) $. The maximal feedback occurs for $\delta_{cp} = \kappa$ because of the phase delay between the pump field and the scattered cavity field.

For a fixed value of $J_\perp$ the effective fermionic Hamiltonian can be diagonalized \cite{CarrNersesyan2006,RouxPoilblanc2007,JaefariFradkin2012,HuegelParedes2013,TokunoGeorges2014}, $H_F=\sum_{k,\sigma= \pm} E_\sigma(k) \gamma^\dagger_{\sigma,k}\gamma_{\sigma,k} $ with $E_\pm(k)=-2J_{\parallel} \cos(\varphi/2)\cos(kd_\parallel)\pm \sqrt{J_\perp^2+4J_\parallel^2\sin^2(\varphi/2)\sin^2(kd_\parallel)}$, where $d_\parallel = \lambda_y/2$ is the lattice spacing along the legs. The details of the Bogoliubov transformation are given in the supplemental material \cite{supp}.

\begin{figure}
\includegraphics[width=0.95\linewidth]{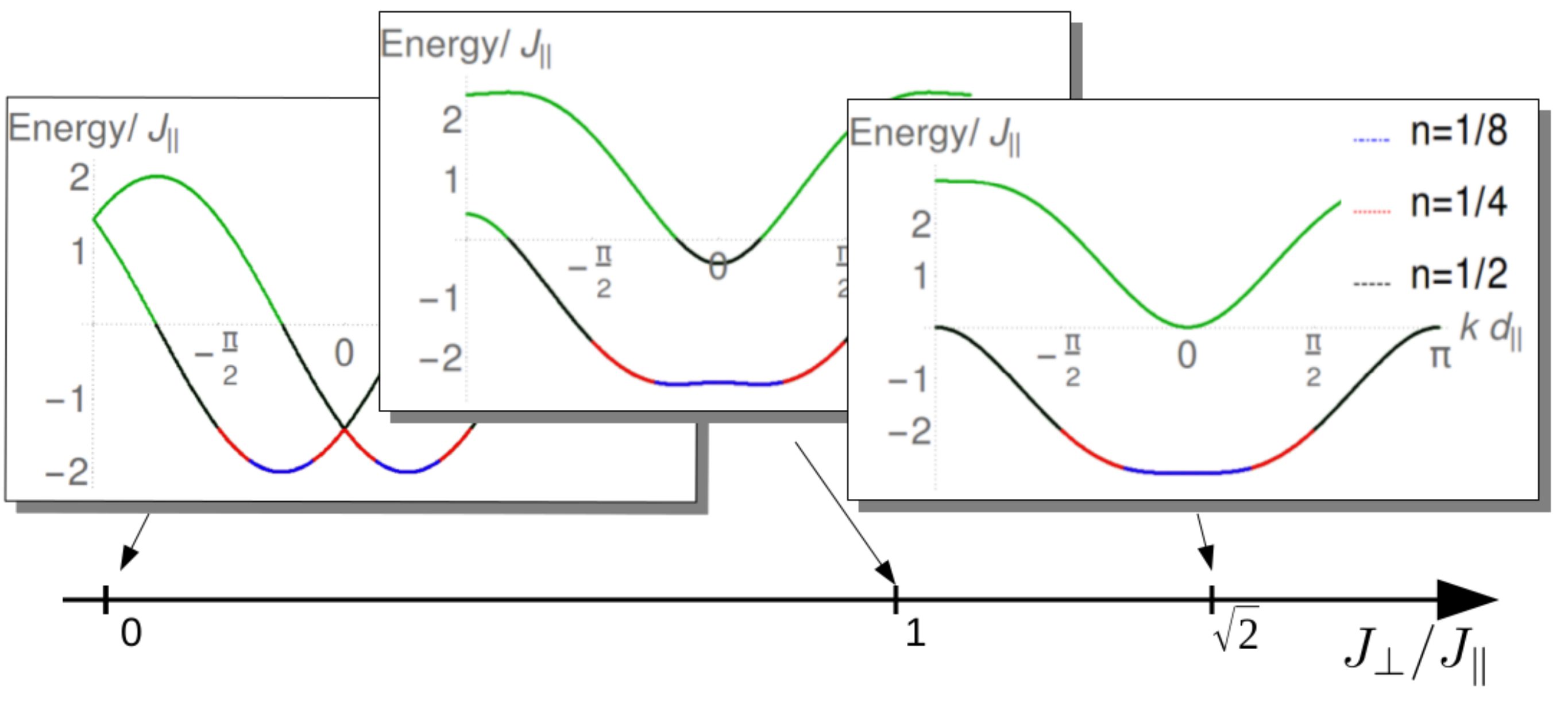}
\caption{\label{fig:energy_comb}(color online) Sketch of the band structure $E_\pm$ for different values of $J_\perp$ at $\varphi=\pi/2$. For $J_\perp<\sqrt{2} J_\parallel$ the lower energy band has two minima $Q_\pm$ and depending on the filling two or four Fermi points occur. At $J_\perp=\sqrt{2} J_\parallel$ the energy bands separate. For  $J_\perp>\sqrt{2}J_\parallel$, the lower energy band has one minimum at $kd_\parallel=0$ and a gap arises between the bands. The colored regions mark the Fermi sea for the filling $n=1/2, 1/4, 1/8$, respectively. Note, that these regions partially lie on top of each other.}
\end{figure}

 The energy dispersion for the case $\varphi=\pi/2$ has two distinct regions (Fig.~\ref{fig:energy_comb}). For $J_\perp<\sqrt{2}J_\parallel$ the two bands are overlapping and the lowest band has two minima at $Q_\pm=\pm 1/d_\parallel \;\arccos\left(\sqrt{\left(\frac{J_\perp} {2J_\parallel}\right)+\frac{1}{2}} \right)$. For $J_\perp>\sqrt{2}J_\parallel$ the two bands are separated by a gap and the lowest band has one minimum at $k=0$. 

One consequence is a transition between a liquid and a band insulator at half filling, $n=N/(2L)=1/2$, where at $J_\perp=\sqrt{2}J_\parallel$  (Fig.~\ref{fig:energy_comb}). Here $N$ is the number of atoms and $L$ the number of rungs on the ladder. At and below the filling $n=1/4$ the occupied fermionic states lie entirely in the lowest energy band $E_-$.

\begin{figure}
\includegraphics[width=0.6\linewidth]{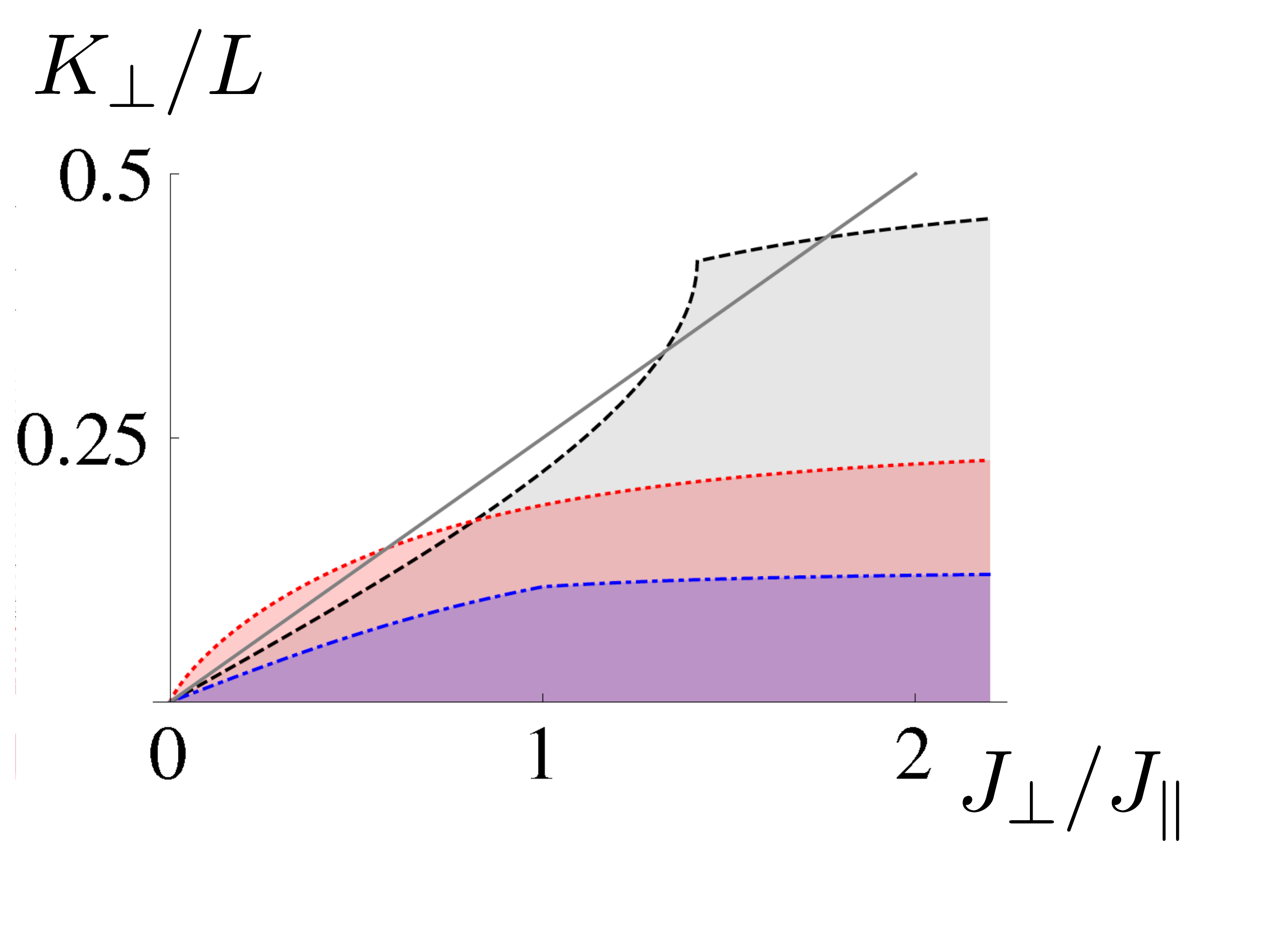}
\caption{\label{fig:hopperp}(color online) The expectation value of the rung tunneling $\aver{K_\perp}/L $ versus the rung tunneling amplitude $J_\perp$ for different fillings $n$. The gray line represents the left-hand side of the self-consistency equation for a fixed value of $AL$. The crossings with the curves indicated by open symbols give the solutions to the self-consistency equation.}
\end{figure}

The influence of the structure of the Fermi sea on the observables can be seen at the example of the rung tunneling $\aver{K_\perp}$ as shown in Fig.~\ref{fig:hopperp}. At large $J_\perp \gg \sqrt{2}J_\parallel$ and filling $n\le 1/2$, the rung tunneling can be approximated by $\aver{K_\perp}/L\approx d_\parallel k_F/(2 \pi)$, where $\pm k_F$ are the two Fermi points which depend linearly on the filling. At half filling the value of the rung tunneling increases slowly for small $J_\perp$ with a concave curvature until at $J_\perp=\sqrt{2} J_\parallel$ it shows a kink and crosses over to the linear behavior. In contrast at $n=1/4$ the rung tunneling shows a very smooth behavior in the entire range. It has a convex curvature at low $J_\perp$ given by a linear rise with a logarithmic correction, i.e.~$\aver{K_\perp}/L\approx \frac{ \left[\log{32} - 2 \log{\left(J_\perp/J_\parallel\right)}\right]}{(2 \sqrt{2} \pi)} J_\perp/J_\parallel+O\left[\left(J_\perp/J_\parallel\right)^3\right]$. At lower filling, e.g.~$n=1/8$, and low $J_\perp$ the behavior is similar to $n=1/4$. However, at low values of $J_\perp$ the expectation value of the directed tunneling $\aver{K_\perp}/L\propto J_\perp/J_\parallel$ and a slight kink occurs at the value $J_{\perp,0}$ where the four Fermi points merge into two Fermi points. The value $J_{\perp,0}$ depends on the filling and is given by $J_{\perp,0}= J_\parallel$ for $n=1/8$ (Fig.~\ref{fig:energy_comb}). 

The gained insights into the dependence of the rung tunneling on $J_\perp$ enables us to find the solutions to the self-consistency equation. In particular, the self-consistency can be interpreted in the form  $J_\perp/A= \aver{K_\perp}$, where the left-hand side stands for a linear curve with slope $1/A$, whereas the right-hand side has the functional dependence of the rung tunneling. The crossings of these two curves give the solutions as indicated in Fig.~\ref{fig:hopperp}.

\begin{figure}
\includegraphics[width=0.8\linewidth]{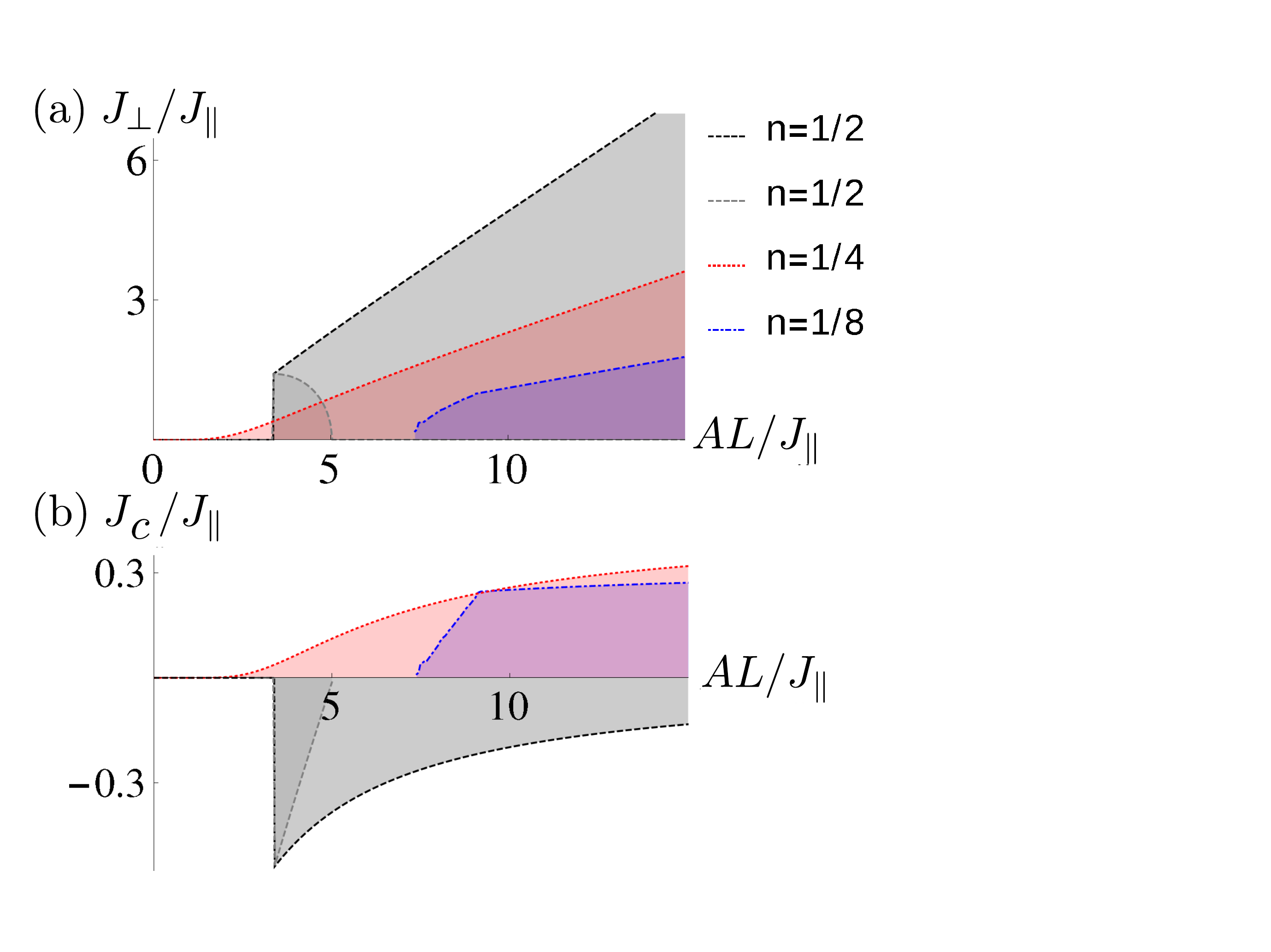}
\caption{\label{fig:alphavsA}(color online) (a) The solutions $J_\perp$ of the self-consistency equation versus the pump strength $AL$ for different fillings $n$. Since the expectation value of the photon operator $\alpha$ is proportional to $J_\perp$, the curves represent also the dependence of the cavity field amplitude on the pump strength. (b) The induced chiral current $J_c$ versus the pump strength for different fillings. For  $n=1/2$ in both quantities a clear jump is present at a critical pump strength. The black and gray dashed lines show the two different solutions of the self-consistency equation in the case of half filling.}
\end{figure}

At half filling up to the critical pump strength of $A_{cr}L= 4\pi J_\parallel/[\sqrt{2}\mathcal{K}(-1)+\mathcal{K}(1/2)]\approx 3.39J_\parallel$ no non-zero solution of the self-consistency equation exists due to the concave curvature of the rung tunneling.  Here, $\mathcal{K}$ denotes the elliptic integral of the first kind. In contrast, for values larger than $A_{cr}$ two solutions exist over a certain range (compare the crossings indicated in Fig.~\ref{fig:hopperp}). The first solution corresponds to values $J_\perp\ge \sqrt{2}J_\parallel$ and exists for all values of $A\ge A_{cr}$.  The second solution corresponds to values $J_\perp<\sqrt{2}J_\parallel$. This solution only exists in an intermediate regime of pump strengths. 
For both cases, the cavity field amplitude jumps to a finite value above the critical pump strength and the fermionic atoms are subjected to an artificial magnetic field (Fig.~\ref{fig:alphavsA}a). For the first solution the expectation value of the cavity field grows monotonically and  depends approximately linearly on the pump strength, i.e.~$\alpha\approx \frac{\tilde{\Omega}L}{\delta_{cp} -i \kappa} \frac{d_\parallel k_F}{(2\pi)}$ at large pump strength. 

The fermionic atoms form a band insulating state which carries a chiral edge current which is defined as $J_c=\sum_j \left(j_{j,0}-j_{j,1}\right) $, where $j_{j,m}=-iJ_\parallel \left( c^\dagger_{\j,m} c_{j+1,m}-\mathrm{H.c.}\right)$ is the local current on leg $m$. In contrast, for the second solution the expectation value of the cavity field decreases with increasing pump strength and the fermionic state corresponds to a liquid which carries a chiral edge current \cite{footnote}. 
In both cases, a sudden onset of the chiral edge current $J_c$ above the critical pump strength $A_{cr}$ occurs (Fig.~\ref{fig:alphavsA}b).  At the jump the chiral current acquires its maximal value and decreases monotonically with increasing pump strength. 

In contrast, at quarter filling, $n=1/4$, this abrupt self-organization changes into a smooth exponentially suppressed activation and the induced fermionic phase corresponds to a chiral liquid. The self-consistency has for all pump strengths a non-zero solution which exhibits a slow initial rise, i.e.~$J_\perp =4\sqrt{2}J_\parallel e^{-2\sqrt{2} \pi J_\parallel/(AL)}$. At larger pump strengths it turns into the linear solution $J_\perp\approx AL d_\parallel k_F/(2\pi)$. Since the Fermi-momenta are lower than in the case of half filling, the reduced slope leads to a slower growth of the tunneling amplitude $J_\perp$. As a consequence, the chiral current exhibits a slow rise with increasing pump strength (Fig.~\ref{fig:alphavsA} b).

 At even lower filling, $n=1/8$, only above a critical value of the pump strength  $A_{cr}L\approx \sqrt{2} \pi J_\parallel/\log\left[\cot(\pi/16)\tan(3\pi/16)\right]\approx 7.3J_\parallel$ a non-vanishing solution of the self-consistency equation exists (Fig.~\ref{fig:alphavsA} a). Above this critical value $A_{cr}$ a slow rise of $J_\perp$ occurs and a self-organization of the combined cavity-atom system into a chiral liquid takes place. The chiral edge current shows two different regimes of increase which are determined by the presence of two or four Fermi points and are connected by a slight kink (Fig.~\ref{fig:alphavsA} b). 

 We briefly outline a possible experimental realization, using the fermionic isotope $^{40}\mathrm{K}$ and cavity parameters $\kappa = \unit[2 \pi \times 4.5]{kHz}$ and $g_0 = \unit[2\pi \times 0.76]{MHz}$ \cite{WolkeHemmerich2012}. The lattice geometry is chosen such that $J_\parallel = 2 \pi \times \unit[100]{Hz}$, $\phi_\parallel \approx 1$, $\phi_\perp \gtrsim 0.05$ and  $\Delta/\hbar = 2 \pi \times \unit[23]{kHz}$ which is sufficiently large compared to the cavity line width. The pump wavelength $\lambda_p = \lambda_x$ is taken to be red-detuned by $\unit[1]{nm}$ from the $\mathrm{D}_2$-line at $\unit[767]{nm}$ and the cavity resonance such that $\delta_{cp} \approx 4 \kappa$. Assuming an effective length of the ladder of $L=10^5$ (in the experiment several parallel ladder structures might be realized) the critical pump strength in the case of half filling, $A_\mathrm{cr} L \approx 3 J_\parallel$, is reached for a pump Rabi frequency of $\Omega^\mathrm{cr}_p=2 \pi \times \unit[100]{MHz}$ which corresponds to an AC Stark shift of $2.4 E_r$, where $E_r =\frac{\hbar^2 k_x^2}{2m}$ and $k_x = 2 \pi/\lambda_x$. Due to off-resonant electronic excitations, the pump light then causes per atom one spontaneous scattering event every $\unit[0.7]{s}$ which is sufficiently large compared to the tunneling time $\hbar/J_\parallel$. Note that, since the effective coupling strength $A$ scales with $1/\delta_{ep}^2$, the spontaneous scattering due to pump light can only be lowered by increasing either $L$, $\phi_\perp$, the cavity Purcell factor $\eta$, or by decreasing $J_\parallel$.

Beyond the mean-field description, an effective dissipative dynamics with the jump operator $K_\perp$ with rate $\Gamma \sim \frac{\kappa \tilde{\Omega}^2}{\delta_{cp}^2+ \kappa^2}$ could drive the atomic system away from the ground state determined by $H_F$ into a steady-state which is a dynamical equilibrium between driving and damping \cite{RitschEsslinger2013}. For the single-pump configuration considered above, this could (at long time-scales) lead -- similar to optical pumping -- to a transfer of the atomic population into the right leg of the ladders (see Fig.~\ref{fig:setup}). This effect could be avoided by adding a second running-wave pump laser field along the $y$-direction which together with a second cavity mode (separated from the first cavity mode by twice the free spectral range) drives Raman transitions along the rungs into the opposite direction. 

The demonstrated self-organization of chiral insulators and chiral liquids in optical cavities could enable a dynamic control and non-destructive detection of such phases. In contrast to previous realizations, here, the artificial magnetic field is generated by the feedback of the cavity mode. The quantum nature of the extended cavity field could lead to an effective (dissipative) long-range interaction between the fermions which could accelerate and stabilize the formation of the correlations. A generalization of the discussed setup to a two-dimensional system \cite{AidelsburgerBloch2013,MiyakeKetterle2013} and to interacting fermionic atoms can be envisaged and will be the subject of further investigations. 

We acknowledge fruitful discussions with M. Fleischhauer, M. K\"ohl, H. Monien, H. Ritsch, and W.~Zwerger and support from the DFG (INST 217/752-1 FUGG).  
\begin{appendix}
\section{Bogolioubov transformation}
The Bogoliubov transformation is determined by the coefficients 

$u_k=\sqrt{\frac{1}{2}\left( 1- \frac{\sin(\varphi/2)\sin(kd_\parallel)}{\sqrt{\left(\frac{J_\perp}{2J_\parallel}\right)^2+\sin^2(\varphi/2)\sin^2(kd_\parallel)}}\right)}$

and

$v_k=\sqrt{\frac{1}{2}\left( 1+ \frac{\sin(\varphi/2)\sin(kd_\parallel)}{\sqrt{\left(\frac{J_\perp}{2J_\parallel}\right)^2+\sin^2(\varphi/2)\sin^2(kd_\parallel)}}\right)}$.

\section{Coefficients from the tight-binding approximation}
In the limit of sufficiently strong optical lattice potentials, the field operators can be expanded into the corresponding Wannier functions $w_y$ along the leg of the ladder and the maximally localized Wannier functions $w_{m}$ of the double well superlattice, where $m=0,1$ labels the right or left well. 
To obtain the effective coupling strength which determines the amplitude of the tunneling process along the cavity direction we can separate the contributions along the $x$- and $y$-direction. 
Along the $y$-direction the following overlap integral is of importance 
\begin{eqnarray}
\nonumber
\phi_\parallel&=& \int \textrm{d}y e^{-ik_p y}  w_y(y) w^*_y(y).
\end{eqnarray}
which is typically in the appropriate geometries close to one. Other contributions to the effective tunneling process along the $x$-direction from overlap intergrals along the $y$-direction can be much smaller and are therefore neglected.

From the $x$-direction two different contributions can play an important role depending on the exact geometry of the optical lattices \cite{JakschZoller2003, AidelsburgerBloch2013,MiyakeKetterle2013,AidelsburgerGoldman2014}. 
The first one is given by the direct overlap integral between the Wannier functions of the two wells and the cavity mode function $\cos\left(k_c x\right)$:
\begin{eqnarray}
\label{eq:phi01}
\phi^{0,1} _\perp&=& \int \textrm{d}x \cos\left(k_c x\right) w_1(x-d_\perp)w^*_0(x).
\end{eqnarray}
 Here $d_{\perp}$ is the lattice spacing along the rungs of the ladder.

The second contribution is given by the time-periodic modulation of the energy offset between the sites of the double well \cite{AidelsburgerGoldman2014}. Its amplitude is related to the on-site overlap integrals
\begin{equation}
\phi^{m,m}_\perp= \int \textrm{d}x \cos\left(k_c (x+m d_\perp) \right) w_m(x)w^*_m(x).
\nonumber
\end{equation}
Since in a double well potential with an energy offset, the values $\phi^{0,0}_\perp$ and $\phi^{1,1}_\perp$ are distinct, the coupling to the cavity mode induces a time-periodic modulation of the energy offset. Such a modulation leads to an effective tunneling along $x$-direction with its amplitude proportional to $\phi^{0,0}_\perp - \phi^{1,1}_\perp$ to first order in the inverse of the oscillation frequency. 

In the main part of the article we use an effective parameter $\phi_\perp$ which includes the two effects. The estimates given are minimal values which are determined from the overlap integral of the Wannier functions (Eq.~\ref{eq:phi01}). 
\end{appendix}


\begin{thebibliography}{34}
\bibitem{footnote}
 \newblock \emph{\bibinfo{journal}{Note that since the system is not in equilibrium the decision which of the solutions it taken, might depend on the initial configuration. A time-dependent solution of the equations would be required which goes beyond the scope of the current work.}}
\end{thebibliography}
\end{document}